%% file: main.tex
\documentclass[conference]{IEEEtran}
\usepackage{fancyhdr}
\IEEEoverridecommandlockouts
\usepackage{microtype}
\usepackage{cite}
\usepackage{amsmath,amssymb,amsfonts}
\usepackage{algorithmic, algorithm}
\usepackage{graphicx}
\usepackage{textcomp}
\usepackage{xcolor}
\usepackage{multirow}
\usepackage{subfigure}
\usepackage{mathtools}
\usepackage{makecell}
\usepackage{caption}
\usepackage{bm}

\DeclarePairedDelimiter\abs{\lvert}{\rvert}

\newtheorem{theorem}{Theorem}
\newtheorem{definition}{Definition}

\newtheorem{lemma}{Lemma}

\def\BibTeX{{\rm B\kern-.05em{\sc i\kern-.025em b}\kern-.08em
    T\kern-.1667em\lower.7ex\hbox{E}\kern-.125emX}}
\begin{document}

\title{Error-controlled Progressive Retrieval of Scientific Data under Derivable Quantities of Interest}

\author{\IEEEauthorblockN{Xuan Wu}
\IEEEauthorblockA{
\textit{University of Kentucky}\\
Lexington, USA \\
xuan.wu@uky.edu}\\
\IEEEauthorblockN{Qing Liu}
\IEEEauthorblockA{
\textit{New Jersey Institute of Technology}\\
Newark, USA \\
qing.liu@njit.edu}\\
\and
\IEEEauthorblockN{Qian Gong}
\IEEEauthorblockA{
\textit{Oak Ridge National Laboratory}\\
Oak Ridge, USA \\
gongq@ornl.gov}\\
\IEEEauthorblockN{Norbert Podhorszki}
\IEEEauthorblockA{
\textit{Oak Ridge National Laboratory}\\
Oak Ridge, USA \\
pnorbert@ornl.gov}\\
\IEEEauthorblockN{Scott Klasky}
\IEEEauthorblockA{\textit{Oak Ridge National Laboratory} \\
Oak Ridge, USA \\
klasky@ornl.gov}
\and
\IEEEauthorblockN{Jieyang Chen}
\IEEEauthorblockA{
\textit{University of Alabama at Birmingham}\\
Birmingham, USA \\
jchen3@uab.edu}\\
\IEEEauthorblockN{Xin Liang\IEEEauthorrefmark{1}\thanks{* Corresponding author: Xin Liang, Department of Computer Science, University of Kentucky, Lexington, KY 40506.}}
\IEEEauthorblockA{
\textit{University of Kentucky}\\
Lexington, USA \\
xliang@uky.edu}\\
}

\maketitle
\thispagestyle{fancy}
\lhead{}
\rhead{}
\chead{}
\lfoot{\footnotesize{
SC24, November 17-22, 2024, Atlanta, Georgia, USA
\newline 979-8-3503-5291-7/24/\$31.00 ©2024 IEEE}}
\rfoot{}
\cfoot{}
\renewcommand{\headrulewidth}{0pt}
\renewcommand{\footrulewidth}{0pt}

\begin{abstract}
The unprecedented amount of scientific data has introduced heavy pressure on the current data storage and transmission systems. 
Progressive compression has been proposed to mitigate this problem, which offers data access with on-demand precision. 
However, existing approaches only consider precision control on primary data, leaving uncertainties on the quantities of interest (QoIs) derived from it. 
In this work, we present a progressive data retrieval framework with guaranteed error control on derivable QoIs. Our contributions are three-fold. (1) We carefully derive the theories to strictly control QoI errors during progressive retrieval. Our theory is generic and can be applied to any QoIs that can be composited by the basis of derivable QoIs proved in the paper. (2) We design and develop a generic progressive retrieval framework based on the proposed theories, and optimize it by exploring feasible progressive representations. (3) We evaluate our framework using five real-world datasets with a diverse set of QoIs. Experiments demonstrate that our framework can faithfully respect any user-specified QoI error bounds in the evaluated applications. This leads to over $2.02\times$ performance gain in data transfer tasks compared to transferring the primary data while guaranteeing a QoI error that is less than 1E-5. 
\end{abstract}


\input{introduction}
\input{related}
\input{formulation}
\input{method}

\input{implementation}

\input{evaluation}
\input{conclusion}

\section*{Acknowledgments}
This research was supported by the Exascale Computing Project CODAR, SIRIUS-2 ASCR research project, the Laboratory Directed Research and Development Program of Oak Ridge National Laboratory (ORNL), and the Scientific Discovery through Advanced Computing (SciDAC) program, specifically the RAPIDS-2 SciDAC institute. 
It was also supported by the National Science Foundation under Grant OAC-2330367, OAC-2311756, OAC-2311757, OAC-2313122, and OIA-2327266. 
We would like to thank the University of Kentucky Center for Computational Sciences and Information Technology Services Research Computing for its support and use of the Lipscomb Compute Cluster, Morgan Compute Cluster, and associated research computing resources.

\newpage

\bibliographystyle{IEEEtran}
\bibliography{reference}


\end{document}

%% file: introduction.tex
\section{Introduction}

The arrival of the first generation of exascale machines and the continuous upgrading of experimental and observational facilities have presented a huge strain on storage, I/O, and networking due to unprecedented data volume and velocities. Because of the limited spacing at high-end parallel file systems (PFS), most of these data must be moved to lower-tier storages, such as tapes, right after generation. Future analyses, which require retrieving data from a central repository and moving across wide area networks, must consider the cost of data retrieval and movement. Recently, lossy compression methods~\cite{lakshminarasimhan2013isabela, sz17, lindstrom2006fast, lindstrom2014fixed, ainsworth2018multilevel} have been developed to tackle the I/O and storage bottleneck as they demonstrate greater compressibility than lossless compressors on floating-point scientific data. Since most simulation and experimental devices inherently involve uncertainty and variability, data can be reduced, provided the loss of accuracy is under prescribed bounds. 


The current leading error-controlled lossy scientific data compressors, including MGARD~\cite{ainsworth2018multilevel, ainsworth2019multilevel, ainsworth2019qoi}, SZ~\cite{sz17,sz18,zhao2021optimizing}, and ZFP~\cite{lindstrom2014fixed}, to name a few, carry mathematically proved theories, which guarantee that errors in the reconstructed data to stay below user-prescribed error bounds. However, most of these compressors only allow to prescribe a single error bound, assuming a ``one-size'' accuracy will fit all subsequent data explorations. 
In contrast, the reduced data may be used for various downstream analyses that are either known or unknown upon data generation. 
To ensure the fidelity of scientific discovery, users have too often conservatively chosen error bounds that cater to the most pessimistic use cases. 
Such over-preservation during data compression, unfortunately, will lead to great expense in data retrieval when faced with a diversity of analyses and use-cases of varying requirements on data fidelity. 

Data refactoring and progressive retrieval pose a potential solution to combat the diverse requests on data fidelity at lower data movement cost \cite{bhatia2022amm, wallace1992jpeg, christopoulos2000jpeg2000,clyne2012progressive,hoang2018study,liang2021error,magri2023general}. Notably, MGARD and ZFP have separately developed the progressive reconstruction feature \cite{liang2021error}, based on multi-level methods (for progressive resolution with MGARD) and bit-plane encoding (for progressive precision with both compressors). They allow data to be archived at nearly full accuracy and retrieved on an as-needed basis, often at reduced resolution and/or precision, for faster data transmission and computations. The progressive reconstruction also allows data to be incrementally recomposed to higher fidelity when more data components become available.   

Despite the potential advantages of progressive retrieval, the gap between the errors in primary data and derived quantities of interest (QoIs) should not be overlooked~\cite{gong2021maintaining}. Obeying the error bounds for QoIs is challenging as the relation between the primary data and the QoI can be highly nonlinear~\cite{lee2022error}. Blindly refining the approximation of the data during progressive retrieval leads to under- or over-estimations, which may not produce correct outcomes in the downstream analyses. Motivated by the disconnection in error control objectives, several works have recently started to explore the preservation of QoIs for a few specific analytic tasks~\cite{liang2020toward, liang2022, jiao2022toward, yan2023toposz} or region-of-interest (RoI) during data compression~\cite{gong2023spatiotemporally}. Nevertheless, directly applying existing QoI-preservation techniques used for compression to progressive retrieval is non-trivial for several reasons: 
\begin{itemize}
    \item QoI-preserving compressors that can handle a broad range of analytic functions are required to prescribe point-wise varied error bounds, whereas the compressors capable of performing progressive retrieval are based on bitplane encoding or multi-level techniques, featuring globally uniform error bounds.
    \item The original values of QoIs are a prerequisite for most QoI-preserving compressors, which are usually computed prior to data compression. Such ground truth values, however, are unattainable with progressive retrieval unless data is recomposed to full fidelity.
    \item These aforementioned difficulties can be further exacerbated when retrieval is targeted at preserving the multivariate and composite QoIs that involve multiple data fields and analytic functions.
\end{itemize}


In this paper, we propose a generic framework to progressively retrieve scientific data with strict error control on \textit{derivable QoIs}. 
We define \textit{derivable QoIs} as downstream quantities that can be explicitly composited by a set of basis functions, including polynomials, square root, and radical functions, along with their combinations through additive, multiplicative, divisive operations, and other functional compositions (see Definition~\ref{def:qoi_uni} and~\ref{def:qoi_multi}). 
The combination of the above basis function and operations will cover a broad range of physical properties, such as kinetic energy, momentum, and magnitude of velocity, that are commonly used in real-world scientific applications. 
Since the derived errors in QoIs vary across data space, we target the $L^\infty$ bound as it measures the extreme case, and the preservation of $L^\infty$ error will automatically ensure the satisfaction of the point-wise error bound.  
In addition, we propose theories to estimate the errors of derivable QoIs based on the errors of primary data, as the ground truth of QoI values cannot be obtained during progressive retrieval, and use the proposed estimators to guide the process of data refinement. 
We further explore and investigate the efficiency of different progressive methods using our framework.
The key contributions are summarized as follows.
\begin{itemize}
    \item We carefully derive the theory to enable QoI error estimation on progressive representations, which can incrementally refine the data approximation until the estimated errors in QoIs are derived from the recomposed data to satisfy user-prescribed bounds. This theory can be generalized to arbitrary error-controlled progressive compressors and offers error control to a broad range of derivable QoIs provided that they can be composed by the basis functions and operations covered in this paper.
    \item We develop a generic progressive retrieval framework capable of QoI error control during progressive retrieval based on our theory. 
    We further integrate three general progressive methods into our framework and explore their efficiency. To this end, we revise the decomposition method in PMGARD~\cite{liang2021error} to enable stable and efficient QoI error control. 
    \item We perform a comprehensive evaluation using scientific data from four real-world applications and one case study with a computational fluid dynamics (CFD) code from Generic Electric (GE). Specifically, we evaluate our framework using different progressive representations and a diverse set of QoIs. Experimental results demonstrate that the proposed method provides strict error control in known QoIs, and this yields over $2.02\times$ performance in data retrieval from remote storage systems via Globus. 
\end{itemize}

The rest of the paper is organized as follows. In Section~\ref{sec:related}, we discuss the related works. In Section~\ref{sec:overview}, we formulate the research problem and present an overview of the compression framework. In Section~\ref{sec:theory}, we introduce the theories to enable QoI error control in progressive formats, which serves as the foundation for the proposed work. In Section~\ref{sec:impl}, we describe the implementation of the proposed framework along with the optimizations. Section~\ref{sec:evaluation} demonstrates the evaluation results with real-world datasets and a case study with GE. In Section~\ref{sec:conclusion}, we conclude the research with a vision for future work.

%% file: related.tex
\section{Related works}\label{sec:related}
In this section, we review the lossy compression and progressive retrieval work derived from the former in the context of scientific data defined on Cartesian grids. For works on tree structure, adaptive meshes, and unstructured data, we refer the readers to \cite{bhatia2022amm,li2018data,hoang2023progressive,said1996new}. 

Data compression is a direct way to mitigate the I/O and storage pressure, which has been studied in the scientific computing community for years. 
Traditional lossless compression techniques~\cite{ziv1977universal, gzip, zstd} achieve only a modest reduction for floating-point scientific data~\cite{son2014data}, which falls far from the desires of exascale computing. Conventional lossy compressors, such as JPEG/JPEG2000~\cite{wallace1992jpeg, christopoulos2000jpeg2000}, while ubiquitous in image transmission, have rarely been used by scientific datasets as they cannot bound errors incurred by compression. 
Therefore, we limit our review to error-controlled scientific compressors. 

The most widely reported error-controlled lossy compressors fall into two broad categories: prediction-based and transform-based.  
Prediction-based compressors such as ISABELA~\cite{lakshminarasimhan2013isabela}, SZ~\cite{sz16, sz17, sz18, zhao2020significantly, liang2018efficient}, and QoZ~\cite{liu2022dynamic} rely on varied predictors, such as spline interpolation or polynomial fitting, to decorrelate the data, whereas transform-based ones such as ZFP \cite{lindstrom2014fixed} and TTHRESH \cite{ballester2019tthresh} employ existing or customized transforms to eliminate redundancy.  
Coefficients after decorrelation/transform may be quantized into integers and then losslessly compressed through entropy or embedded encoding approaches to reduce the size. Notably, scientific lossy compressors carry mathematical theories for quantization and encoding, which ensure the maximal error between the original and reconstructed data is less than a user-specified error bound. Recently, several compressors even advanced the error control onto downstream QoIs that are derived from the reconstructed data~\cite{ainsworth2019qoi, liang2020toward, liang2022, jiao2022toward, yan2023toposz, lee2022error,banerjee2022algorithmic}.

MGARD~\cite{ainsworth2018multilevel, ainsworth2019multilevel, ainsworth2019qoi} derives a \textit{norm} based on the finite element analysis and wavelet theories, applying it to tighter the error bounds such that the most pessimistic QoI cases can be satisfied. 
Due to the complexity of mathematical derivation, MGARD's current QoI-control theory is only applicable to linear QoIs, which limits its use cases. 
A variation of SZ has also been proposed in~\cite{jiao2022toward}, which relies on a pre-evaluation of target QoIs and derivating point-wise compression error bounds such that QoI values computed from the reconstructed data will satisfy user-prescribed error bounds. 
The post-processing-based QoI-preserved techniques~\cite{lee2022error,banerjee2022algorithmic} iteratively update the reduced approximation until the derived QoI errors stay below prescribed bounds. Nevertheless, similar to the QoI preservation work with SZ, the post-processing technique requires knowing the original QoI values and is only applicable to univariate QoIs. 
Several additional compression methods have been developed to reduce the data while preserving topological features such as critical points~\cite{liang2020toward, liang2022} and contour trees~\cite{yan2023toposz}, but they do not generalize to other QoIs.


The most prominent downside of lossy compressors is that the ``one-size-fits-all'' error prescription strategy is prone to underestimating or wasting resources when faced with diverse post-processing tasks. 
In contrast, progressive compression and retrieval allow for dynamic adjustment of the transmitted data size based on requested fidelity and support incremental recomposition to obtain finer data representations without starting from scratch. 
The progressive approaches can be generally categorized into progression in resolution and progression in precision. The most well-known approaches in the former category include Fourier and discrete cosine transform~\cite{zhang2019efficient}, wavelets pyramid~\cite{li2019vapor,li2023lossy}, multi-level decomposition~\cite{ainsworth2019multilevel}, and rank decomposition \cite{ballard2020tuckermpi,ballester2019tthresh}, where data representations in coarser resolution are obtained by retrieving only a subset of coefficients. In comparison, progression in precision is often achieved through encoding the bit planes \cite{liang2021error,li2023lossy,lindstrom2014fixed}, or iteratively compressing the residues with progressively decreased error bounds \cite{magri2023general}. 
With bit-plane encoding, the precision-based progressive retrieval will be performed among all coefficients, starting from the most to least significant bit. 
With progressively decreased error bounds, the compression procedure will generate multiple snapshots with different precision for retrieval. Progression in precision can provide more fine-grained retrieval compared to progression in resolution. We also notice that some progressive techniques, such as the PMGARD~\cite{liang2021error}, support progression in both categories. Specifically, PMGARD combines the orthogonal decomposition method in MGARD with bitplane encoding to provide guaranteed error control on primary data. 

Despite the potential to fulfill data requests of arbitrary precision, none of the existing progressive compression techniques provide error control on downstream QoIs.
In this work, we bridge the gap by developing a generic framework to determine the proper amount of data to retrieve in progressive formats to meet user-specified QoI tolerances, which is expected to significantly reduce the retrieval size and thus improve data movement performance. 
To the best of our knowledge, this is the first attempt to tackle QoI preservation during progressive retrieval.




%% file: formulation.tex
\section{Overview}\label{sec:overview}
We formulate our research problem in this section, followed by an overview of the proposed framework. 
The notations used in the paper are summarized in Table~\ref{tab:notations}.

\begin{table}[t]
\vspace{2mm}
\centering
\caption{Notations}
\label{tab:notations}
\footnotesize
\resizebox{\columnwidth}{!}{%
\begin{tabular}{|c|l|}
\hline
\thead{Symbol} & \thead{Description} \\ \hline
$n_e$                            & Number of data points. \\ \hline
$n_v$                            & Number of variables. \\ \hline
$n_q$                            & Number of target QoIs. \\ \hline
$n_s$                            & Number of progressive segments. \\ \hline
$\tau$                           & Error tolerance on QoIs. \\ \hline
$x, x'$                          & Single scalar values. \\ \hline
$s$                              & Data fragments produced by progressive compression. \\ \hline
$\epsilon$                       & Error bound on the primary data. \\ \hline
$\xi$                            & Real error in the primary data (upper bounded by $\epsilon$). \\ \hline
$\mathbf{x}, \mathbf{x^{\prime}}, \bm{\epsilon}$ & Vectors of $x, x^\prime, \epsilon$ in multivariate cases. \\ \hline
$x_i, x^{\prime}_i, \epsilon_i$  & The $i$-th element in $\mathbf{x}, \mathbf{x^{\prime}}, \bm{\epsilon}$. \\ \hline
$f$                              & Univariate QoI that applies to data on a single field. \\ \hline
$g$                              & Multivariate QoI that applies to data on multiple fields. \\ \hline
$\Delta(f, x, \epsilon)$         & Upper bound of QoI error in $f$ at $x$ with error bound $\epsilon$. \\ \hline
$\Delta(g, \bm{x}, \bm{\epsilon})$ & Upper bound of QoI error in $g$ at $\bm{x}$ with error bound $\bm{\epsilon}$. \\ \hline
$\lvert \cdot \rvert$            & Operator of getting absolute value. \\ \hline
$\{\cdot_i\}$                    & An array of the referred element. \\ \hline
\end{tabular}
}
\end{table}

\begin{figure*}[t]
	\includegraphics[width=\textwidth]{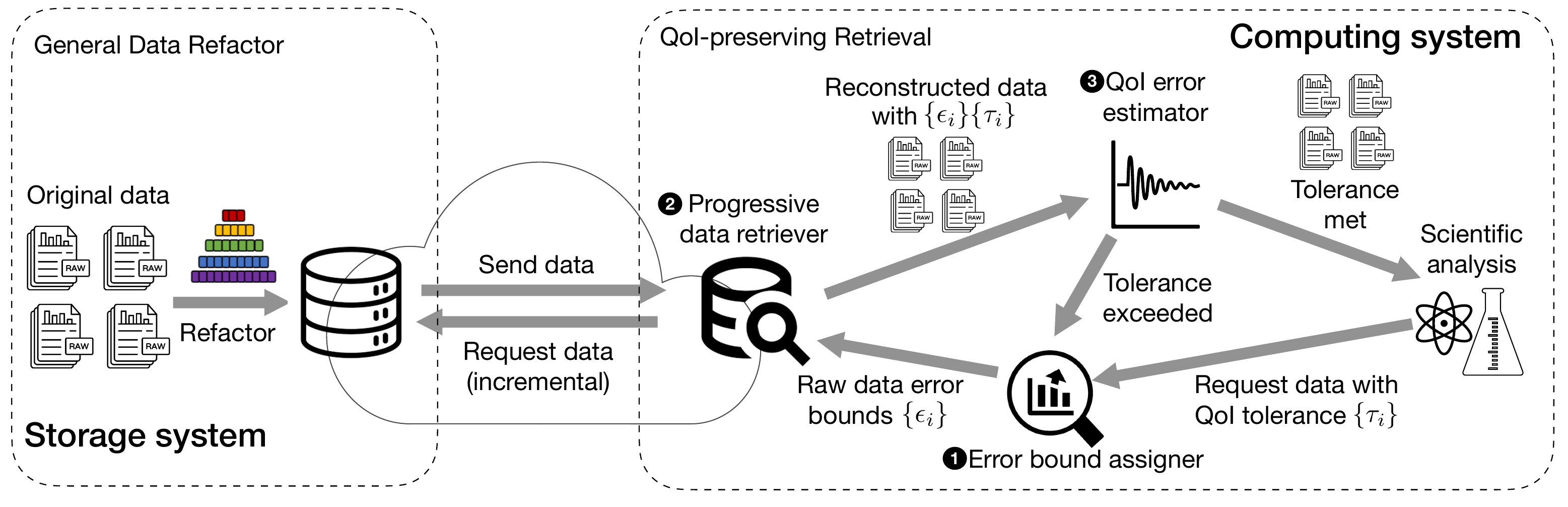}
	\centering
	\caption{Workflow of the proposed QoI-preserving framework with three key modules. We assume that data is refactored and stored in storage systems when generated, and our framework is able to progressively retrieve data from storage while guaranteeing user-specified QoI error bounds. This is extremely useful when data movement becomes the performance bottleneck, which is the case when data is located in secondary or remote storage systems.}
\label{fig:overview}
\end{figure*}

\subsection{Problem formulation}
Our progressive retrieval framework is designed to extract only the ``necessary'' amount of progressive fragments and guarantee that the user-prescribed error tolerance in QoIs derived from the reconstructed data is met. 
The capability of estimating the errors in QoIs is crucial for the trustability of the reconstructed data, and it can accelerate the process of reading data from low-tier storage or remote central repository by minimizing the data volume. 
Below, we define the requested functionalities in progressive compressors and the \textit{derivable} QoIs targeted in this paper.

\begin{definition}\label{def:progressive}
An error-controlled progressive compression method shall be able to (1) refactor the original data $\{x'_1, \cdots, x'_{n_e}\}$ into progressive fragments $\{s_1, \cdots, s_{n_s}\}$ for archiving, where $n_e$ is the number of original data points and $n_s$ is the number of progressive segments;
(2) reconstruct data $\{x_1, \cdots, x_{n_e}\}$ from $\{s_1, \cdots, s_j\}$ such that $\max_i\abs{x'_i - x_i} < \epsilon_j$, where $\epsilon_j$ is the prescribed error bound when recomposing data using the first $j$ segments.
\end{definition}

\begin{definition}\label{def:qoi_uni}
A univariate QoI is a univariate function $f: \mathbb{R} \to \mathbb{R}$ that maps a scalar value to a quantity. The derivable univariate QoIs include a family of QoIs that can be formulated by the set of basis functions defined in Table \ref{tab:qois}.
\end{definition}

\begin{definition}\label{def:qoi_multi}
A multivariate QoI is a multivariate function $g: \mathbb{R}^n \to \mathbb{R}$ that maps a vector to a quantity. The derivable multivariate QoIs include a family of QoIs that can be composited using the derivable univariate QoIs and operations of derivable multivariate QoIs defined in Table \ref{tab:qois}.
\end{definition}

\begin{table}[ht]
\centering
\caption{Bases of Derivable QoIs}
\label{tab:qois}
\footnotesize
\resizebox{\columnwidth}{!}{%
\begin{tabular}{|c|c|l|c|}
\hline
\thead{Name} & \thead{Category} & \thead{Formula} & \thead{Example in GE} \\ \hline
Polynomials & univariate & $f(x) = \sum{a_ix^i}$ & \eqref{eq:1} \eqref{eq:5}, \eqref{eq:6}\\ 
\hline
Square root & univariate & $f(x) = \sqrt{x}$ & \eqref{eq:1}, \eqref{eq:3}, \eqref{eq:6}  \\ 
\hline
Radical & univariate & $f(x) = 1/(x + c)$ & \eqref{eq:6}  \\ 
\hline
Addition & multivariate & $g(x_1, \cdots, x_n) = \sum x_i$ & \eqref{eq:1} \\ 
\hline
Multiplication & multivariate & $g(x_1, \cdots, x_n) = \Pi x_i$ & \eqref{eq:5}, \eqref{eq:6}  \\ 
\hline
Division & multivariate & $g(x_1, x_2) = x_1 / x_2$ & \eqref{eq:2}, \eqref{eq:4}  \\ 
\hline
Composition & both & $(f_1 \circ f_2)(x) = f_1(f_2(x))$ & \eqref{eq:1} -- \eqref{eq:6}  \\ 
\hline
\end{tabular}
}
\end{table}

Despite the fact that there are only seven families of QoIs listed above, their compositions cover a broad range of functions commonly used in scientific studies.  
Below, we showcase how to break down the user-requested QoIs into the basis of derivable QoIs using the non-proprietary data generated by a Computational Fluid Dynamics (CFD) code from GE. 
The CFD simulation produces velocity $V_x$, $V_y$, $V_z$, pressure $P$, and density $D$ on a set of unstructured meshes. We linearize the data into a one-dimensional array, choosing six QoIs used in the posthoc analyses with the detailed equations listed below, as they will be used in our experimental evaluation later on:
\begin{align}
    \label{eq:1}
    V_{total} &= \sqrt{\bm{V_x}^2 + \bm{V_y}^2 + \bm{V_z}^2},\\ 
    \label{eq:2}
    T &= \frac{\bm{P}}{\bm{D}*R},\\
    \label{eq:3}
    C &= \sqrt{\gamma*R*\bm{T}}, \\
    \label{eq:4}
    Mach &= \frac{\bm{V_{total}}}{\bm{C}}, \\ 
    \label{eq:5}    
    PT &= \bm{P} (1+\frac{\gamma}{2}*\bm{Mach}*\bm{Mach})^{mi}, \\
    \label{eq:6}
    \mu &= \mu_r (\frac{\bm{T}}{T_r})^{1.5} \frac{T_r + S}{\bm{T} + S}.
\end{align}
Here, $R = 287.1$, $\gamma = 1.4$, $mi = 3.5$, $\mu_r = 1.716e-5$, $T_r = 273.15$, and $S=110.4$ are different constant values, and the input and intermediate variables are bolded. 
Using $PT$ defined in Equation~\eqref{eq:5} as an example, the formula can be decomposed into the multiplication of $P$ and $(1+\frac{\gamma}{2}*Mach*Mach)^{mi}$, where the latter is a composition of the square root function and a polynomial of $Mach$.

\subsection{Design overview}
We illustrate the workflow of the proposed QoI-preserving progressive data retrieval framework in Fig.~\ref{fig:overview}.
Our key contribution lies in the retrieval procedure, which iteratively refines the reduced approximation till the estimate errors in QoIs drop below user-prescribed bounds. 
Specifically, we develop (1) a QoI error estimator that provides an upper bound of the errors in QoI given the reconstructed data and its $L^\infty$ error bounds, and (2) a primary data (PD) error assigner which estimates and prescribes the bounds used for the next round of data retrieval.
The progressive compression procedure can be performed using the existing error-controlled progressive compression frameworks~\cite{lindstrom2014fixed, liang2021error, magri2023general}, where data is refactored and compressed into multi-precision fragments.

The overall data retrieval pipeline can be summarized as follows. Firstly, an analytic task requests a set of QoIs and the desired error tolerance.
This request is processed by the PD error-bound assigner (module 1), which gauges the error bounds on each primary data field used by the first round of retrieval. 
Such error bounds will be sent to a progressive retriever (module 2), which extracts progressive segments from the most to the least significant until the errors in the errors in the reconstructed data reach below the requested bounds. 
Data will be incrementally recomposed using the newly arrived progressive segments and then fed into
the QoI error estimator, along with the error bounds used during retrieval, to estimate the upper bounds of QoI errors under the current data representations (module 3). 
If the estimated QoI errors are less than the requested tolerances, the data is provisioned for the analyses; otherwise, the current data representations, along with the derived QoI errors, will be forwarded to the PD error bound assigner to estimate the error bounds used in the next round of data retrieval. 
The pipeline repeats these steps till the targeted QoI error bounds are reached, or a full-fidelity data representation is retrieved.  
Due to their progressive nature, only incremental portions of the data need to be retrieved in the later requests, which promises high efficiency in managing the data movement from storage systems to applications.

\subsection{Quality assessment}
We leverage the widely used rate-distortion curves~\cite{liang2021error, liu2022optimizing, liu2024high} to evaluate the efficiency of our approach. 
The X-axis in the curve is bitrate, which represents the average number of bits in the compressed format. 
It is analogous to the compression ratio in single-snapshot compression, and can be computed by the retrieved data size divided by the number of elements.
We use relative QoI errors as our distortion metric for the Y-axis, which is computed by the maximal absolute error of QoI divided by its respective value range. 
An easy way to compare multiple rate-distortion curves is to fix either the X value or Y value: in the former case, one can compare the errors of different approaches based on the same retrieved data size; in the latter case, one can compare the size of retrieved data under the same quality.  

%% file: method.tex
\section{Theoretical foundation}\label{sec:theory}


In this section, we introduce the theoretical foundation of the proposed work. The data retrieval is designed to meet the error bounds prescribed on QoIs.  
Below, we describe how to estimate the errors in QoI using the reconstructed data and its $L^\infty$ error bound during data retrieval. 
Having such error estimation is critical as we need to iteratively update and examine the QoI errors during data retrieval. Specifically, we derive the upper error bounds for the bases of derivable QoIs (shown in Table \ref{tab:qois}) and discuss their combinations in univariate, multivariate, and composite cases.

The following subsections start with the definition of QoI errors for each case, then theorems and proofs for different types of QoI functions. Please refer to Table \ref{tab:notations} for the notations of the symbols used in our theorems and proofs. 
Notably, throughout the derivations, we assume the original data $x'$ and the reconstructed data $x$ to satisfy a $L^\infty$ error bound, as will be detailed below.
 
\subsection{Univariate QoIs}
\begin{definition}\label{def:preservable_uni}
Given a data value $x$ and a $L^\infty$ error bound $\epsilon$ used during progressive retrieval, we define $\Delta(f, x, \epsilon)$ as the supremum of QoI error under a univariate QoI $f(x)$: $\Delta(f, x, \epsilon) = \sup_{\abs{x' - x} \leq \epsilon}{\abs{f(x') - f(x)}}$.
\end{definition}

Here, we assume the original value is $x'$ will satisfy the error bound constraint $\abs{x' - x} \leq \epsilon$, then we have $\abs{f(x^*) - f(x)}_\infty \leq \sup_{\abs{x' - x} \leq \epsilon}{\abs{f(x') - f(x)}}_\infty = \Delta(f, x, \epsilon)$.
Note that $\Delta(f, x, \epsilon)$ only relies on the reconstructed data and the error bound used during data retrieval. 
Below, we present the theorems used for estimating $\Delta(f, x, \epsilon)$ given several univariate QoI functions.

\begin{theorem}\label{theorem:power}
[Polynomials] An upper bound of $\Delta(f, x, \epsilon)$ for function $f(x) = x^n$ can be written as $\Delta(f, x, \epsilon) \leq \sum_{i=1}^{n}{C_n^i \abs{x}^{n-i}\epsilon^i}$, where $C_n^i=\frac{n!}{(n-i)!i!}$ is the combination formula.
\end{theorem}
\begin{IEEEproof}
$\abs{f(x') - f(x)}= \abs{(x + \xi)^n - x^n} = \abs{\sum_{i=0}^n \xi^ix^{(n-i)} - x^n} = \abs{\sum_{i=1}^{n}{C_n^i x^{n-i}\xi^i}} \leq \sum_{i=1}^{n}{C_n^i \abs{x}^{n-i}\abs{\xi^i}} \leq \sum_{i=1}^{n}{C_n^i \abs{x}^{n-i}\epsilon^i} $.

\end{IEEEproof}


\begin{theorem}\label{theorem:sqrt}
[Square Root] An upper bound for function $f(x) = \sqrt{x}$ can be written as $\Delta(f, x, \epsilon) \leq {\epsilon} / (\sqrt{\max(x - \epsilon, 0)} + \sqrt{x}).$
\end{theorem}
\begin{IEEEproof}
Since the negative $x-\epsilon$ can be replaced by 0, $\sqrt{x'} = \sqrt{x+\xi} \geq \sqrt{\max(0, x-\epsilon)}$. Then
$\abs{f(x') - f(x)} = \abs{\sqrt{x + \xi} - \sqrt{x}} = \abs{{\xi} / (\sqrt{x + \xi} + \sqrt{x})} \leq {\epsilon} / (\sqrt{\max(x - \epsilon, 0)} + \sqrt{x})$.
\end{IEEEproof}

\begin{theorem}\label{theorem:radical}
[Radical] An upper bound for radical function $f(x) = 1 / ({x + c})$ can be written as $\Delta(f, x, \epsilon) \leq {\epsilon} / \{\min(\abs{x+c-\epsilon}, \abs{x+c+\epsilon})\abs{x+c}\}$, when $x + c\neq 0$ and $\epsilon < \abs{x+c}$.
\end{theorem}
\begin{IEEEproof}
$\abs{f(x') - f(x)} = \abs{{1} / ({x+\xi+c}) - {1} / ({x+c})} = \abs{{\xi} / ((x+\xi+c)(x+c))} $. Since $\epsilon < \abs{x+c}$, we have $\abs{x+\xi+c} \geq \min(\abs{x+c-\epsilon}, \abs{x+c+\epsilon})$. Therefore, $\abs{f(x') - f(x)} \leq {\epsilon} / \{\min(\abs{x+c-\epsilon}, \abs{x+c+\epsilon})\abs{x+c}\}$.
\end{IEEEproof}

Note that Theorem \ref{theorem:radical} does not apply to the case of $\epsilon > \abs{x+c}$, as it may lead to an infinitesimal value of $\abs{x+\xi+c}$, causing the errors in QoI unable to be bounded. 
Such a case can be avoided by only choosing $\epsilon < \abs{x+c}$ during data retrieval.

\subsection{Multivariate QoIs}

\begin{definition}\label{def:preservable_uni}
Given a vector $\bm{x} = (x_1, \cdots, x_n)^T$ and a $L^\infty$ error bound vector $\bm{\epsilon}$ used during data retrieval, we define $\Delta(g, \bm{x}, \bm{\epsilon})$ as the supremum of QoI error under a multivariate QoI $g$: $\Delta(g, \bm{x}, \bm{\epsilon}) = \sup_{\abs{\bm{x'} - \bm{x}}  \leqq \bm{\epsilon}}{\abs{g(\bm{x'}) - g(\bm{x})}}$, where $\leqq$ indicates that the \textit{less and equal to} relationship is applied for every element in the input vector.
\end{definition}

Assume $\xi_i = x_i' - x_i$ and $x'_i \in [x_i - \epsilon_i, x_i + \epsilon_i]$ following the error bound constraints, then we have the following theorems.

\begin{theorem}\label{theorem:add_var}
[Addition] An upper bound for weighted summation function $g(\bm{x}) = \sum_{i=1}^{n}{a_i x_i}$ is $\Delta(g, \bm{x}, \bm{\epsilon}) \leq \sum_{i=1}^{n}{ \abs{a_i}\epsilon_i}$.
\end{theorem}
\begin{IEEEproof}
$\abs{g(\bm{x'}) - g(\bm{x})} = \abs{\sum_{i=1}^{n}{a_i\xi_i}} \leq \sum_{i=1}^{n}\abs{{ a_i}}\abs{\xi_i} \leq \sum_{i=1}^{n}{ \abs{a_i}\epsilon_i}$.
\end{IEEEproof}

\begin{theorem}\label{theorem:mult_var}
[Multiplication] An upper bound for multiplication function $g(x_1, x_2) = x_1 x_2$ can be written as $\Delta(g, \bm{x}, \bm{\epsilon}) \leq \abs{x_1}\epsilon_2 + \abs{x_2}\epsilon_1 + \epsilon_1\epsilon_2$.
\end{theorem}
\begin{IEEEproof}
$\abs{g(\bm{x'}) - g(\bm{x})} = \abs{(x_1 + \xi_1)(x_2 + \xi_2) - x_1x_2} = \abs{x_1 \xi_2 + x_2\xi_1 + \xi_1 \xi_2} \leq \abs{x_1}\abs{\xi_2} + \abs{x_2}\abs{\xi_1} + \abs{\xi_1\xi_2} \leq \abs{x_1}\epsilon_2 + \abs{x_2}\epsilon_1 + \epsilon_1\epsilon_2$.
\end{IEEEproof}

\begin{theorem}\label{theorem:div_var}
[Division] An upper bound for division function $g(x_1, x_2) = {x_1} / {x_2}$ can be written as $\Delta(g, \bm{x}, \bm{\epsilon}) \leq (\abs{x_1}\epsilon_2 + \abs{x_2}\epsilon_1) / \{\abs{x_2}\min{(\abs{x_2 - \epsilon_2}, \abs{x_2 + \epsilon_2})}\}$ when $\epsilon < \abs{x_2}$.
\end{theorem}
\begin{IEEEproof}
$\abs{g(\bm{x'}) - g(\bm{x})} = \abs{({x_1 + \xi_1}) / ({x_2 + \xi_2}) - {x_1} / {x_2}} = \abs{(x_2\xi_1-x_1\xi_2) / \{(x_2+\xi_2)x_2\}} \leq (\abs{x_2}\abs{\xi_1} + \abs{x_1}\abs{\xi_2}) / (\abs{x_2 + \xi_2} \abs{x_2}) \leq ({\abs{x_1}\epsilon_2 + \abs{x_2}\epsilon_1}) / \{{\abs{x_2}\min{(\abs{x_2 - \epsilon_2}, \abs{x_2 + \epsilon_2})}}\}
$.
\end{IEEEproof}

\subsection{Composite QoIs}
This subsection applies the theories of addition, scalar multiplication, and composition for the QoI functions presented in the previous subsection to broaden the range of QoIs that can be preserved during retrieval. 
Specifically, we have the following theorems for composite QoIs.

\begin{theorem}\label{theorem:add}
[Additive] $\Delta(f, x, \epsilon)$ satisfies the additive property $\Delta(f_1 + f_2, x, \epsilon) \leq \Delta(f_1, x, \epsilon) + \Delta(f_2, x, \epsilon)$.
\end{theorem}
\begin{IEEEproof}
$\Delta(f_1 + f_2, x, \epsilon) = \abs{f_1(x+\xi) + f_2(x+\xi) - f_1(x) - f_2(x)} \leq \abs{f_1(x+\xi) - f_1(x)} + \abs{f_2(x+\xi) - f_2(x)} = \Delta(f_1, x, \epsilon) + \Delta(f_2, x, \epsilon)$. 
\end{IEEEproof}

\begin{theorem}\label{theorem:mult}
[Multiplicative] $\Delta(f, x, \epsilon)$ satisfies the multiplicative property $\Delta(af, x, \epsilon) = a\Delta(f, x, \epsilon)$ for any constant $a\neq 0$.
\end{theorem}
\begin{IEEEproof}
$\Delta(af, x, \epsilon) = \abs{af(x+\xi) - af(x)} = a\abs{f(x+\xi) - f(x)} = a\Delta(f, x, \epsilon)$. 
\end{IEEEproof}

\begin{theorem}\label{theorem:composite}
[Composition] $\Delta(f, x, \epsilon)$ satisfies the composite property $\Delta(f_1 \circ f_2, x, \epsilon) = \Delta(f_1, f_2(x), \Delta(f_2, x, \epsilon))$.
\end{theorem}
\begin{IEEEproof}
Denote $y=f_2(x)$ and $\xi'=f_2(x+\xi) - f_2(x)$, then 
$\xi' \in [f_2(x) - \Delta(f_2, x, \epsilon), f_2(x) + \Delta(f_2, x, \epsilon)]$.
Correspondingly,  
$\Delta(f_1 \circ f_2, x, \epsilon) = \abs{f_1(f_2(x+\xi)) - f_1(f_2(x))} = \abs{f_1(y + \xi') - f_1(y)} = \Delta(f_1, f_2(x), \Delta(f_2, x, \epsilon))$. 
\end{IEEEproof}

Although the above proofs have been conducted on univariate QoIs, the same theorems apply to the additive, multiplicative, and composite operations in multivariate QoIs. 
Additionally, we derive the error estimators for the composition of univariate QoIs and multivariate QoIs and summarize them in the two lemmas below. 
We omit the details of the proofs due to limited space. The proofs are similar to the procedure in Theorem~\ref{theorem:composite}.

\begin{lemma}\label{lem:fg}
Denote $f \circ g$ as the composition of a univariate function $f$ and a multivariate function $g$ such that $f \circ g(\bm{x}) = f(g(\bm{x}))$. We have $\Delta(f \circ g, \bm{x}, \bm{\epsilon}) = \Delta(f, g(\bm{x}), \Delta(g, \bm{x}, \bm{\epsilon})).$
\end{lemma}
\begin{lemma}\label{lem:gf}
Denote $g \circ \{f_1, \cdots, f_n\}$ as an element-wise composition of a multivariate function $g$ and $n$ univariate functions $\{f_1, \cdots, f_n\}$ such that $g \circ \{f_1, \cdots, f_n\}(x_1, \cdots, x_n) = g(f_1(x_1), \cdots, f_n(x_n))$. We have $\Delta(g \circ \{f_1, \cdots, f_n\}, \bm{x}, \bm{\epsilon}) = \Delta(g, \bm{x'}, \Delta(g, \bm{x}, \bm{\epsilon'}))$ where $\bm{x'} = (f_1(x_1), \cdots, f_n(x_n))^T$ and $\bm{\epsilon'} = (\Delta(f_1, x_1, \epsilon_1), \cdots, \Delta(f_n, x_n, \epsilon_n))^T$.
\end{lemma}

These composite theorems and lemmas greatly extend our flexibility, allowing for progressively retrieving and bounding errors in a variety of QoIs. 
For instance, multiplications of multiple variables in the form of $\Pi{x_i}$ can be preserved by iteratively leveraging the multiplication theory (Theorem~\ref{theorem:mult_var}) and composite property (Theorem~\ref{theorem:composite}). Errors in a general polynomial in the form of $\sum{a_ix^i}$ can be upper bounded by applying the additive (Theorem~\ref{theorem:add}), multiplicative (Theorem~\ref{theorem:mult}) properties, and the error estimation of power functions (Theorem~\ref{theorem:power}).

\subsection{Example derivation on GE case study}
Here, we showcase how to estimate the $V_\textrm{total}$ in the GE case study using the proposed methods. 
Let $x_1, x_2, x_3$ denote $V_x$, $V_y$, $V_z$. The $V_\textrm{total}$, as denoted in Equation~\eqref{eq:1}, can be formulated as the composition of a univariate function $f_1(x) = \sqrt{x}$ with the composition of a multivariate function $g_1(x_1, x_2, x_3) = x_1 + x_2 + x_3$ and an univariate function $f_2(x) = x^2$, which yields $V_{total} = f_1(g_1(f_2(x_1), f_2(x_2), f_2(x_3)))$. 
We estimate the upper bound of errors in $f_2$, $g_1$, and $f_1$ sequentially. First, the upper bound of errors in $f_2(x_i)$ ($i=1,2,3$) can be estimated using Theorem~\ref{theorem:power}. 
This forms the error bound vector $\bm{\epsilon_{f2}} = (\Delta(f_2, x_1, \epsilon_1), \Delta(f_2, x_2, \epsilon_2), \Delta(f_2, x_3, \epsilon_3))^T $.
Meanwhile, the value of $f_2(x_i)$ can be computed to obtain the new value vector $\bm{x_{f2}} = (f_2(x_1), f_2(x_2), f_2(x_3))$. 
After that, $\bm{\epsilon_{f2}}$ and $\bm{x_{f2}}$ will be used to compute $\epsilon_{g1} = \Delta(g_1 \circ f_2, \bm{x_{f2}}, \bm{\epsilon_{f2}})$ using Theorem~\ref{theorem:add_var}. 
At last, we will compute $x_{g1} = g_1(f_2(x_1), f_2(x_2), f_2(x_3))$ to derive the final QoI error bound $\Delta(f_1 \circ g_1 \circ f_2, x_{g1}, \epsilon_{g1})$ using Theorem~\ref{theorem:sqrt}.

While using QoIs in GE as a demonstrative example, our theories are extendable to other scientific applications due to the following reasons. 
First, some common QoIs in the paper can be directly used by other applications (e.g., total velocity in climatology and cosmology). Second, the set of basis QoIs can composite diverse and complex functions such as multivariate polynomials and rational functions, which cover a broad range of QoIs, including molar concentration multiplications in combustion. 
Third, our theory can extend to new operators with derivable error control (e.g., isosurface~\cite{jiao2022toward}). 
This demonstrates the genericity of the proposed QoI-preserving theory.  



%% file: implementation.tex
\section{Implementation and Optimization}\label{sec:impl}
In this section, we present the implementation of the QoI-preserving progressive retrieval framework, followed by optimizations and explorations on efficiency. 

\subsection{Algorithm and implementation}
Our pipeline consists of two stages: a \textit{data refactoring} stage, which transforms the original data into multi-precision segments for storage, and a \textit{data retrieval} stage, which fetches and recomposes data till it reaches user-specified QoI tolerances. 
We omit the discussion on progressive refactoring as it's a direct application of the existing techniques. 
Generally speaking, for every data field, the refactoring stage produces a set of multi-precision segments and the corresponding metadata. 

\begin{algorithm}[h!]
\caption{\textsc{General Data Refactor}} \label{alg:refactor} \footnotesize
\renewcommand{\algorithmiccomment}[1]{/*#1*/}
\begin{flushleft}
\textbf{Input}: number of variables $n_v$, all variables $\{v_i\}$\\
\textbf{Output}: refactored multi-precision segments $\{\{s_p\}_i\}$ and metadata $\{m_i\}$
\end{flushleft}
\begin{algorithmic} [1]
\FOR{$j = 1 \to n_v$}
    \STATE $\{s_p\}_i, m_i = \texttt{refactor}(v_i)$
\ENDFOR
\end{algorithmic}
\end{algorithm}

\begin{algorithm}[h!]
\caption{\textsc{QoI-preserved Data Retrieval}} \label{alg:retrieval} \footnotesize
\renewcommand{\algorithmiccomment}[1]{/*#1*/}
\begin{flushleft}
\textbf{Input}: refactored multi-precision segments $\{\{s_p\}_i\}$ and metadata $\{m_i\}$, value range of original variables $\{range_i\}$, requested QoI tolerances $\{\tau_i\}$\\
\textbf{Output}: retrieved data 
\end{flushleft}
\begin{algorithmic} [1]

\FOR{$i = 1 \to n_v$}
    \STATE $v_i \gets \{0, \cdots, 0\}$  \COMMENT{initial all the data fields as zero vector}
\ENDFOR
\FOR{$j = 1 \to n_v$}
    \STATE $\epsilon_j \gets $ \texttt{assign\_eb}$(range_j, \{\tau_i\})$ \COMMENT{assign the initial error bounds}
\ENDFOR
\STATE tolerance\_met $\gets$ \textbf{false}
\WHILE{$!$tolerance\_met}
    \FOR{$i = 1 \to n_v$}
        \STATE $v_i = \texttt{progressive\_construct}(v_i, \{s_p\}_i, m_i, \epsilon_i)$ \COMMENT{construct the $i$-th variable to the target precision $\epsilon_i$}
    \ENDFOR
    \STATE tolerance\_met $\gets$ \textbf{true} \COMMENT{initialize flag as true}
    \STATE $\{\tau'_k\} \gets 0$ \COMMENT{initialize max estimated QoI errors}
    \FOR{$j = 1 \to n_e$}
        \FOR{$k=1 \to n_q$}
            \STATE $\tau' \gets \texttt{estimate\_error}(\{v_i\}_k, \{\epsilon_i\}_k)$ 
            \COMMENT{estimate QoI errors under current representation based on Section~\ref{sec:theory}}
            \IF{$\tau' > \tau_k$}
                \STATE tolerance\_met $\gets$ \textbf{false} \COMMENT{if the errors of at least one QoIs are not met, set flag to false for another iteration}
                \IF{$\tau' > \tau'_k$}
                    \STATE $\tau'_k \gets \tau', ind_k \gets j$ \COMMENT{record position of max error}
                \ENDIF
            \ENDIF
        \ENDFOR
    \ENDFOR
    \FOR{$k = 1 \to n_q$}
        \STATE $\{\epsilon_i\} \gets $ \texttt{reassign\_eb}$(\tau'_k, ind_k, \tau_k, \{v_i\}, \{\epsilon_i\})$ \COMMENT{compute the new error bounds for variables based on data with max QoI errors}
    \ENDFOR
\ENDWHILE
\RETURN $\{v_i\}$
\end{algorithmic}
\end{algorithm}

The proposed QoI-preserved data retrieval pipeline is presented in Algorithm~\ref{alg:retrieval}, assuming that the retrieval starts from scratch. 
Lines 1-6 show the initialization of the progressive data representation and error bounds used in the first round of retrieval.  
The \texttt{while} loop starting from line 7 presents the iterative procedure used for QoI-preserving data retrieval that the data representations are gradually refined and then checked for QoI errors till all QoI tolerances are satisfied or a full-fidelity data representation has been retrieved.
The \texttt{progressive\_construct} function takes the newly retrieved multi-precision segments and recomposes the current data representation to a more accurate approximation. 
Lines 13-23 estimate the errors of QoIs derived from the reconstructed data using the theorems and lemmas discussed in Section \ref{sec:theory}, and compare them against user-requested error tolerance. 
When the requested tolerances are not met, 
we record the maximal estimated errors as well as their corresponding locations in the data space. Such information will be used for optimizing the error bound assigned for the next round of data retrieval (lines 17 - 22).

Regarding the PD error bound used for data retrieval, the initialization and iterative refinement stages adopt separate error assignment algorithms.  
At the initialization (line 5) stage, we adopt the Algorithm~\ref{alg:assign_eb}: when a data field is utilized by multiple QoIs, its error bound will be determined by the minimal relative tolerance among all the requested QoIs involving this field. 
At the iterative refinement (line 26) stage, we adopt a uniform error-tightening strategy detailed in Algorithm~\ref{alg:reassign_eb}, and prioritize the evaluation of the data point that generates the largest QoI errors in the previous evaluation. 
If the estimated QoI errors exceed the tolerance, we reduce the error bounds of all the variables involved in the computation for this QoI by a constant factor $c$ ($c=1.5$ in our implementation), iteratively retrieve the data, then estimate the QoI errors under the new error bounds and the reconstruction. 
Note that we execute the QoI error estimation only on the data point with the largest QoI error at the iterative refinement stage, which decreases the number of required iterations in Algorithm~\ref{alg:retrieval}.

\begin{algorithm}[t]
\caption{\textsc{assign\_eb}} \label{alg:assign_eb} \footnotesize
\renewcommand{\algorithmiccomment}[1]{/*#1*/}
\begin{flushleft}
\textbf{Input}: value range $range$, requested QoI tolerances $\{\tau_i\}$ \\
\textbf{Output}: error bound of current variable
\end{flushleft}
\begin{algorithmic} [1]
\STATE $\epsilon$ $\gets $ 1 \COMMENT{initilize eb to maximal possible relative bound}
\FOR{$j = 1 \to n_q$}
    \IF{the $j$-th QoI involves this variable}
        \STATE $\epsilon$ = \texttt{min}(eb, $\tau_j$)
    \ENDIF
\ENDFOR
\RETURN $\epsilon*range$
\end{algorithmic}
\end{algorithm}

\begin{algorithm}[t]
\caption{\textsc{reassign\_eb}} \label{alg:reassign_eb} \footnotesize
\renewcommand{\algorithmiccomment}[1]{/*#1*/}
\begin{flushleft}
\textbf{Input}: index $k$ with largest QoI error, requested QoI tolerance $\tau_k$, current data $\{v_i\}$, current error bounds $\{\epsilon_i\}$, reduction factor $c=1.5$ \\
\textbf{Output}: new error bound of the current variables
\end{flushleft}
\begin{algorithmic} [1]
\STATE $\tau' \gets \texttt{estimate\_error}(\{v_i\}_k, \{\epsilon_i\}_k)$ 
            \COMMENT{re-estimate QoI errors under potentially updated error bounds; see Section~\ref{sec:theory} for details}
\WHILE{$\tau' > \tau$}
    \FOR{$v_i$ involved in this QoI}
        \STATE $\epsilon_i = \epsilon / c$
    \ENDFOR
    \STATE $\tau' \gets \texttt{estimate\_error}(\{v_i\}_k, \{\epsilon_i\}_k)$ 
            \COMMENT{re-estimate again}
\ENDWHILE
\RETURN $\{\epsilon_i\}$
\end{algorithmic}
\end{algorithm}

We have also implemented a mask-based outlier management method that filters out irregular points that potentially lead to unbounded error estimation. 
Using the CFD simulation data generated by GE as the example again, for nodes with values of $V_x = V_y = V_z = 0$, their decompressed values will be tiny when the error bounds used for retrieval are small.
These close-to-zero values, however, could yield loose upper bounds for $V_\textrm{total}$ (seeing Theorem~\ref{theorem:sqrt}) despite the small real errors. 
Accoordingly, in this example, we will use a bit-map to record the position of any data point with $0$ total velocity, and only refactor data points whose values are non-zero. 

\subsection{Exploration on the best-fit progressive representation}\label{sec:exploration}
Despite the fact that the proposed QoI error-control theories and pipeline can be generalized to any progressive compressors that meet the Definition \ref{def:progressive}, the amount of data retrieved may vary due to the different progressive refactoring and error bounding theories adopted by each compressor. 
In this section, we review and evaluate the pros and cons of three error-controlled progressive compression algorithms. 

\textbf{Error-controlled compression with multiple snapshots: } This type of methods leverages existing error-controlled compressors to compress data using a list of error bounds $\{\epsilon_i\}$, ranging from small to large. 
When an error bound $\epsilon^*$ is requested during progressive retrieval, one can choose a snapshot with a minimal $i$ such that $\epsilon_i < \epsilon^*$. 
Due to the overlapping information among these snapshots, redundancies may be high when multiple precisions are requested during the progressive retrieval.

\textbf{Error-controlled delta compression~\cite{magri2023general}: } This type of methods also reduces data into multiple snapshots of multi-precision segments but eliminates the redundancy across the snapshots by compressing the residues (errors between the original and decompressed data) instead of the original data. 
As a result, it can be more efficient than directly compressing data into multiple snapshots with different error bounds but requires retrieving all first $i$ snapshots ($i$ is the minimal integer such that $\epsilon_i < \epsilon^*$) when the target error bound is $\epsilon^*$. 

\textbf{Progressive compression with bitplane: } This type of methods does not require to pre-set error bounds. Instead, it encodes the data using bitplanes such that the data can be retrieved and recomposed on demand. 
Similar to the error-controlled delta compression, it may not be the most efficient when only a single error bound is requested -- directly compressing data using the requested error bound usually generates the smallest data footprints in such cases. 
PMGARD~\cite{liang2021error} is the leading technique in this kind, but its performance suffers from loose error control and thus may return more precision fragments than needed. 

Below, we evaluate the performance of these three kinds of methods when a set of successively lower relative error bounds (i.e., a series of requests $\{\epsilon_i\}$ such that $\epsilon_{i+1} < \epsilon_{i}$) are requested. 
For the first two categories of progressive compressors, we use $SZ3$ as the underlying error-controlled compressors as it provides the tightest $L^\infty$ error bound and thus could yield larger compression ratios than compressors with looser error bound~\cite{liang2018efficient, tao2019optimizing}. 
From now on, we refer to the integration of progression with multiple-snapshot, progression with delta compression, and SZ3 as \texttt{PSZ3} and \texttt{PSZ3\_delta}, respectively. 
For progression with bitplane, we use PMGARD as the underlying refactor. The evaluations have been conducted using GE's CFD simulation data and their six QoIs (see Table~\ref{tab:data} for details on datasets).

\begin{figure}[ht]
	\includegraphics[width=\columnwidth]{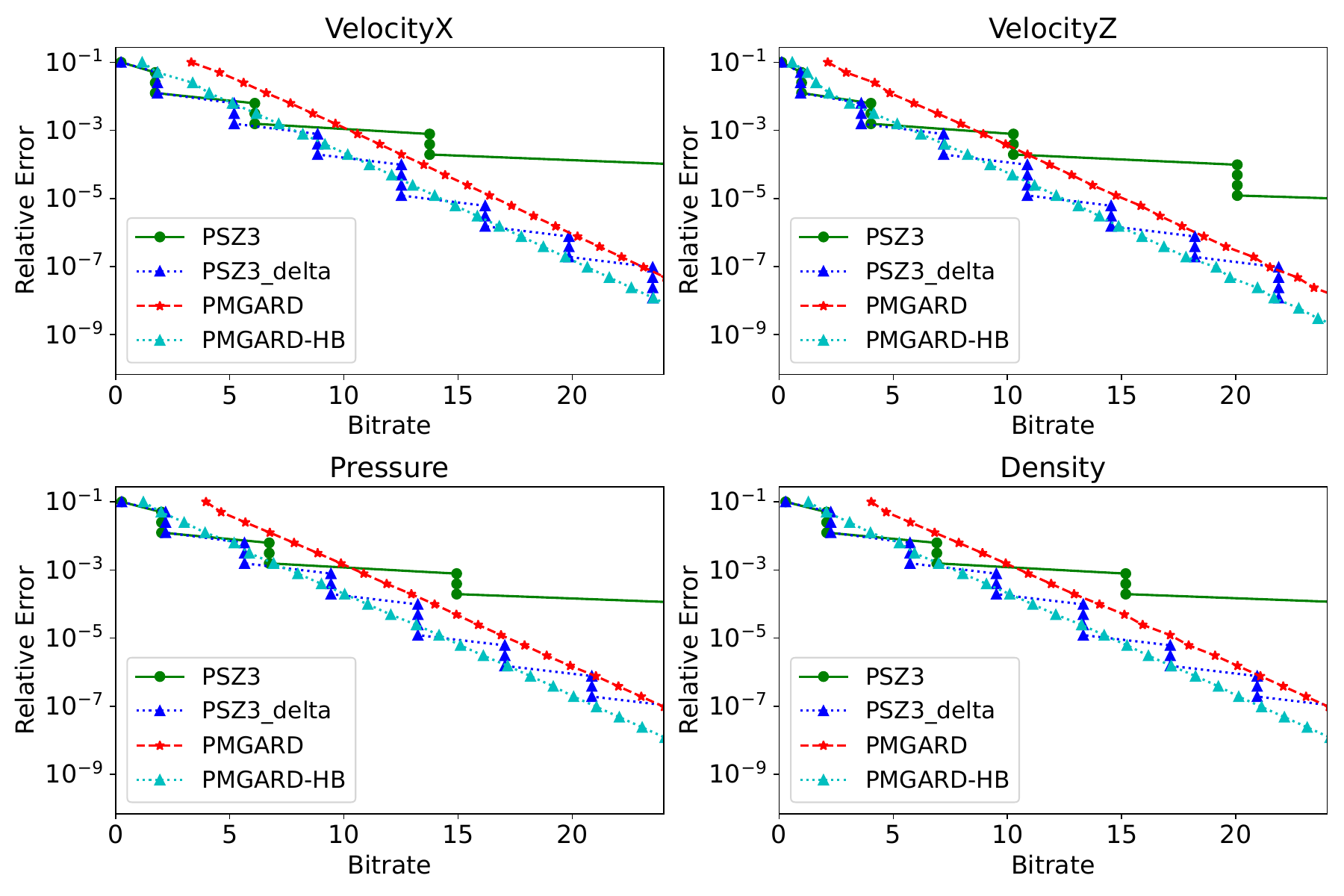}
	\centering
	\caption{Requested error and the resulting bitrate for different error-controlled progressive compressors.}\label{fig:error}
\end{figure}

We evaluated $4$ data fields in Fig.~\ref{fig:error}. 
The trend of VelocityY is similar to those of VelocityX and VelocityZ, and thus omitted. 
We set $\epsilon_i = 10^{-i}$ for $i = 1, 2, \cdots 10$ for both \texttt{PSZ3} and \texttt{PSZ3\_delta} to create multiple snapshots (because this setting has a reasonable trade-off between additional storage cost and retrieval efficiency), and requested errors bounds on primary data as $\{\epsilon'_i\} = 0.1*2^{-i}$ for $i=1, 2, \cdots 20$.
The rate-distortion curves can be interpreted as follows: for any data point $(x, y)$, its value represents the bitrate $x$ (analogous to the percentage of data retrieved) under the requested tolerance $y$. As such, the closer a curve is to the left bottom (left indicates lower retrieve percentage and bottom indicates low error), the better. 
Notably, \texttt{PSZ3} has large bitrates under progressive requests, which is expected due to the redundancy among the multiple snapshots. 
Since \texttt{PSZ3} and \texttt{PSZ3\_delta} rely on pre-set error bounds to produce multi-precision fragments, their bitrate curves exhibit a stair-case pattern: the amount of retrieved data remain constant across several adjacent error bounds and then present a sudden drop as the retrieval moves to the next snapshot.  
This leads to suboptimal results when the desired error bound is slightly lower than one of the pre-set error bounds. 
On the contrary, the trends of bitrate in \texttt{PMGARD} are linear with respect to the requested error bounds. 
Nevertheless, \texttt{PMGARD} constantly generates larger bitrates than those of \texttt{PSZ3-delta}, as the error bounds implemented with the former are looser than the latter. 

\begin{figure}[ht]
	\includegraphics[width=\columnwidth]{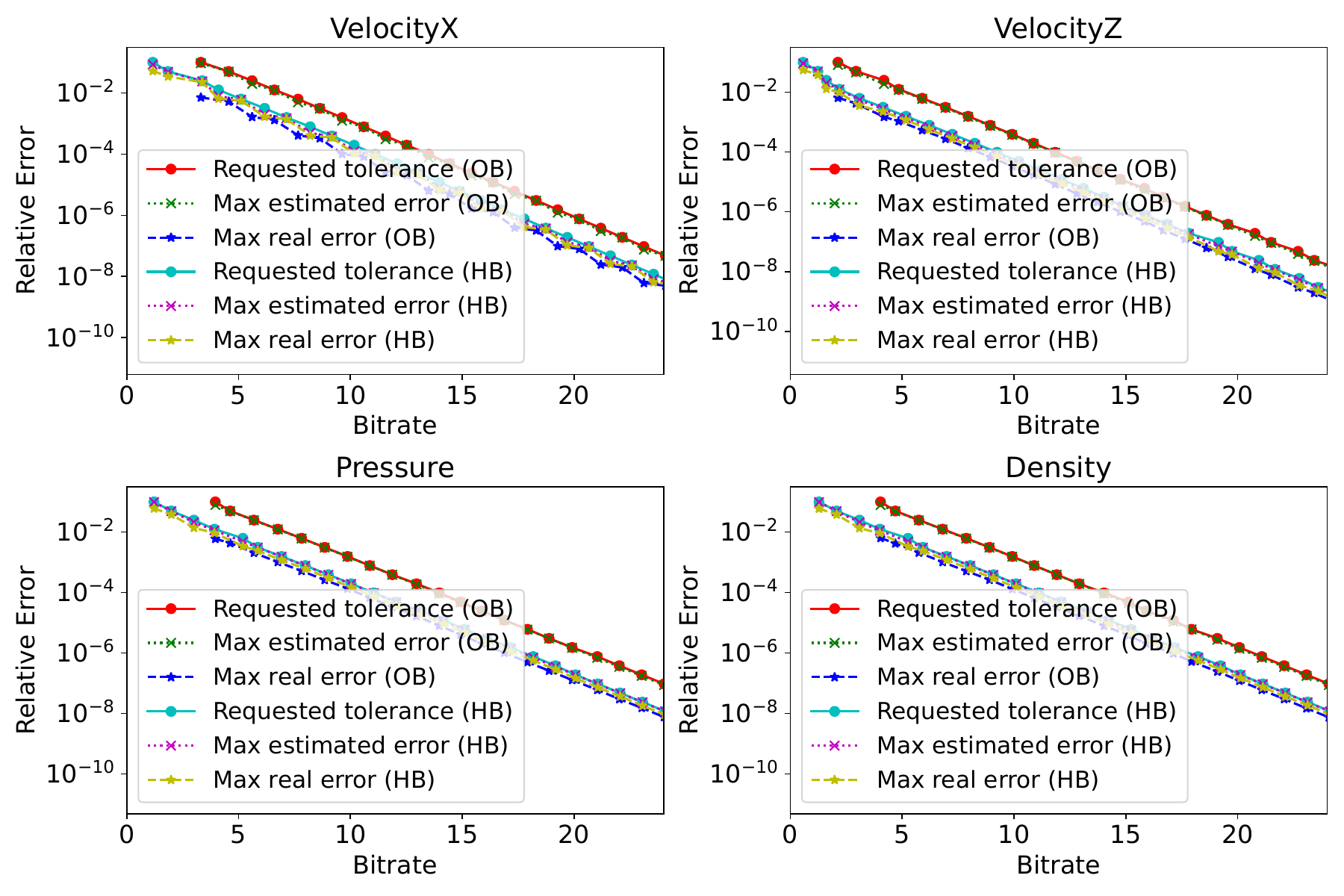}
	\centering
	\caption{Impact of decomposition basis on GE-small data. OB and HB represent \texttt{PMGARD} and \texttt{PMGARD-HB}, respectively.}\label{fig:decomposition}
\end{figure}

For \texttt{PMGARD}, the gap between the requested bounds and real errors is mainly caused by the decomposition algorithm, in particular a $L^2$ projection, which maps low-level coefficient nodes to high-level nodal nodes employed for data decorrelation.  
This decomposition algorithm is derived from MGARD, which is specifically designed to provide optimal error control in $L^2$ norm, whereas can cause over-pessimistic estimation on $L^\infty$. 
This is further validated by the experimental results shown in Fig~\ref{fig:decomposition}, which examines the difference among the requested tolerance, estimated upper bound, and actual error measured after progressive retrieval. It shows that while the estimated errors are close to the requested tolerance, the actual errors are far smaller, causing an over-retrieval problem. 

We propose to reduce the gap by omitting the $L^2$ projection in \texttt{PMGARD}'s decomposition algorithm.  
This could yield two benefits. 
First, without the cross-level intervention, the $L^\infty$ norm can be accurately estimated through a summation of the maximal error bounds across all levels. 
Second, because $L^2$ projection is time-consuming, removing it can accelerate the refactoring and reconstruction process. 
We call the new progressive algorithm without $L^2$ projection \texttt{PMGARD-HB} as it replaces the orthogonal basis in MGARD with a hierarchical basis from a mathematical point of perspective.  
As shown in Fig.~\ref{fig:decomposition}, \texttt{PMGARD-HB} yields much more accurate error estimation than \texttt{PMGARD}, and this leads to constantly lower bitrates in all requested error bounds. 
We further compare \texttt{PMGARD-HB} with \texttt{PSZ3} and \texttt{PSZ3\_delta} in Fig.~\ref{fig:error}, where one can observe the advantages in most bitrates.



%% file: evaluation.tex
\section{evaluation}\label{sec:evaluation}
We evaluate our methods in terms of error control, progressive retrieval efficiency, and performance using five datasets from four real-world applications with respective QoIs.
Specifically, we first validate the error control of the QoI-preserving theory in Section~\ref{sec:theory}, and then demonstrate the efficiency of three progressive methods --\texttt{PSZ3}, \texttt{PSZ3-delta}, and \texttt{PMGARD-HB}-- which are the representatives of three mainstream error-controlled progressive approaches according to the literature. 

\subsection{Experiment setup}
We conduct all our experiments using the Morgan Compute Cluster (MCC)~\cite{mcc} located at the University of Kentucky with 100 Gbps InfiniBand HDR interconnect. 
Each compute node in the system is equipped with 2 AMD EPYC ROME 7702P processors, each with 64 cores and 256 GB memory.
Our benchmark datasets are from multiple computational science domains including CFD, cosmology~\cite{almgren2013nyx}, climate~\cite{hurricane-data}, and combustion~\cite{chen2017s3d}, and their specifications are listed in Table~\ref{tab:data}. 
Note that we use $\{\_\_\}$ in the dimensionality of GE data because its second dimension may have variable sizes. 
We use the first four datasets for sequential evaluation and the last dataset for measuring data transfer performance in a distributed memory environment.   
For QoIs, we use Equation~\eqref{eq:1} --~\eqref{eq:6} for the two GE datasets, and we test total velocity (i.e., $V_{total}$ in Equation~\eqref{eq:1}, also referred to as VTOT in the rest of the evaluation) for NYX and Hurricane data to demonstrate the generalibility. 
For S3D data, the data represent the molar concentration of $8$ species associated with $21$ reactions, and their multiplications generate the intermediate variables to derive the rate of progress. 
In particular, we present $4$ multiplications involved in $2$ reactions. For instance, $x_0, x_1, x_3, x_4, x_5$ used in our evaluation represent species $H_2, O_2, H, O, OH$, respectively, and $x_1x_3$ computes the molar concentration of $O_2$ and $H$ in the reaction $H + O_2 \xrightleftharpoons{} O + OH$.
While we convert single-precision NYX and Hurricane data to double-precision for evaluation with small error bounds, our method directly applies to single-precision floating-point data.

\begin{table}[ht]
{
\footnotesize
\centering
\caption{Datasets and QoIs}
\label{tab:data}
\footnotesize
\resizebox{\columnwidth}{!}{%
\begin{tabular}{|l|c|c|c|c|c|}
\hline
\thead{Dataset} & \thead{Dimensions} & \thead{$n_v$} & \thead{Type} & \thead{Size} & \thead{QoIs}\\ 
\hline
GE-small & $200\times \{\_\_\}$ & 5 & double & 137.96 MB & $Eq.\eqref{eq:1}$ -- $\eqref{eq:6}$ \\
\hline
Hurricane & $100 \times 500 \times 500$ & 3 & double &  572.20 MB & Total velocity\\
\hline
NYX & $512\times 512\times 512$ & 3 & double & 3.00 GB & Total velocity \\
\hline
S3D & $1200\times 334\times 200$ & 8 & double & 4.78 GB 
& \begin{tabular}{@{}c@{}}Molar concentration \\ multiplication\end{tabular} \\
\hline
GE-large & $96 \times \{\_\_\} $ & 5 & double & 7.79 GB &  $Eq.\eqref{eq:1}$ -- $\eqref{eq:6}$ \\
\hline
\end{tabular}
}
}
\end{table}

\subsection{QoI error control}

\begin{figure}[h!]
	\includegraphics[width=\columnwidth]{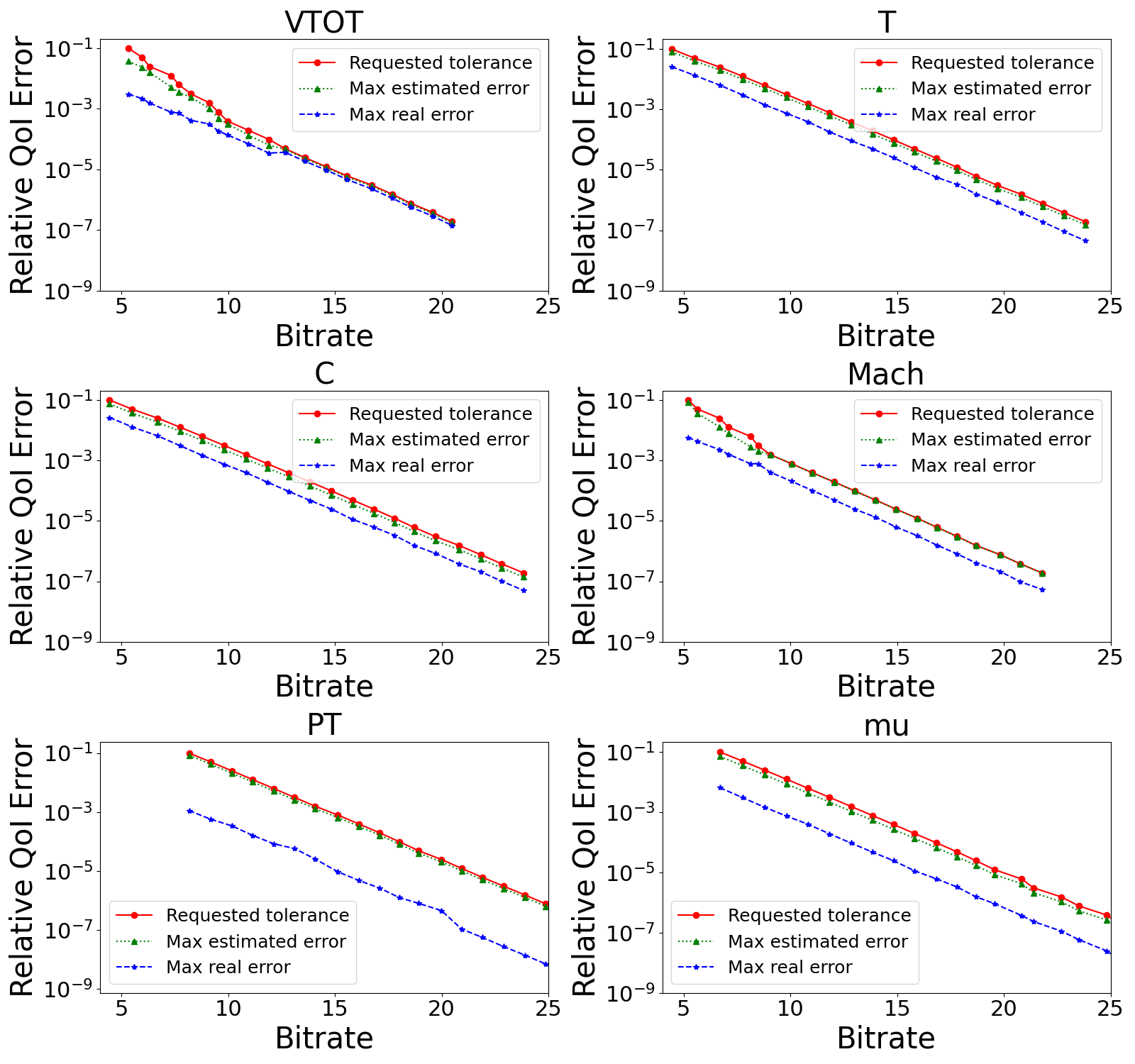}
	\centering
	\caption{Max estimated and max actual QoI errors under given requested QoI errors of \texttt{PMGARD-HB} on GE-small.}\label{fig:qoi-ge}
\end{figure}

We first show that our theory provides guaranteed error control on the derivable QoIs in the evaluated applications.
We use \texttt{PMGARD-HB} for demonstration purposes, and the same functionality can be provided by \texttt{PSZ3} and \texttt{PSZ3-delta} as well. 
In particular, we present the max estimated QoI errors and actual QoI errors of the proposed method under a progressive set of requested QoI errors for GE data in Figs.~\ref{fig:qoi-ge}. 
It is observed that the actual QoI errors in our method are always smaller than the estimated QoI errors, which are usually close but strictly smaller than the requested QoI errors. 
This validates our theory, which always provides an upper bound for the QoI error estimations. 
Different trends are observed for different variables as well. 
For instance, one can see a gap between the max estimated errors and actual errors in total velocity when the bitrate is low.
This is because some decompressed data become close to $0$ when the error bound is high, in which case the estimation of $\sqrt{x}$ generally leads to a loose bound. 
This situation becomes better when the error bound decreases to a certain threshold, which implies the diminishment of near-zero decompressed values. 
Furthermore, one can notice a larger gap between the max estimated errors and actual errors in PT than that in the other QoIs, which is reasonable because the estimation in PT is the most complex and involves more relaxation. 
In addition, the trends in T and C are similar, which is also as expected because their formulas are very similar. 

\begin{figure}[t]
	\includegraphics[width=\columnwidth]{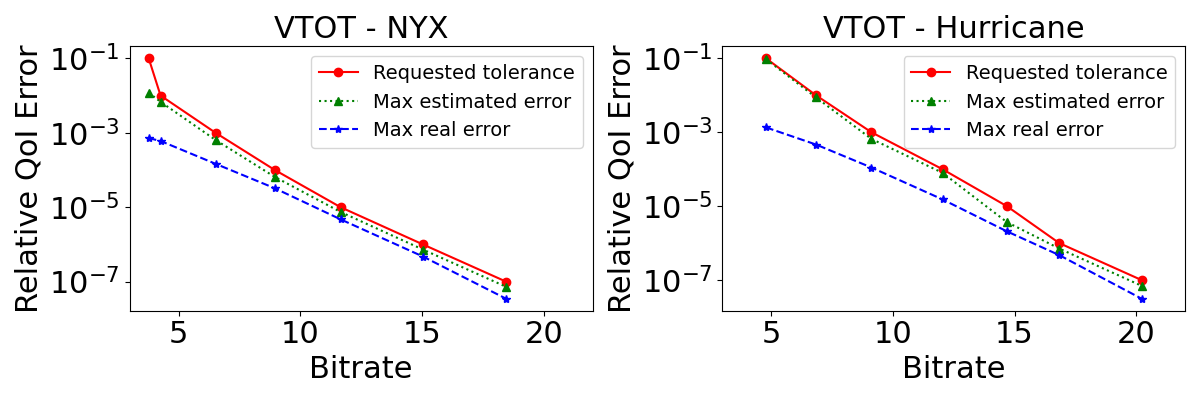}
	\centering
	\caption{Max estimated and max actual QoI errors under given requested QoI errors of \texttt{PMGARD-HB} on NYX and Hurricane.}\label{fig:qoi-nyx-hurricane}
\end{figure}

\begin{figure}[t]
	\includegraphics[width=\columnwidth]{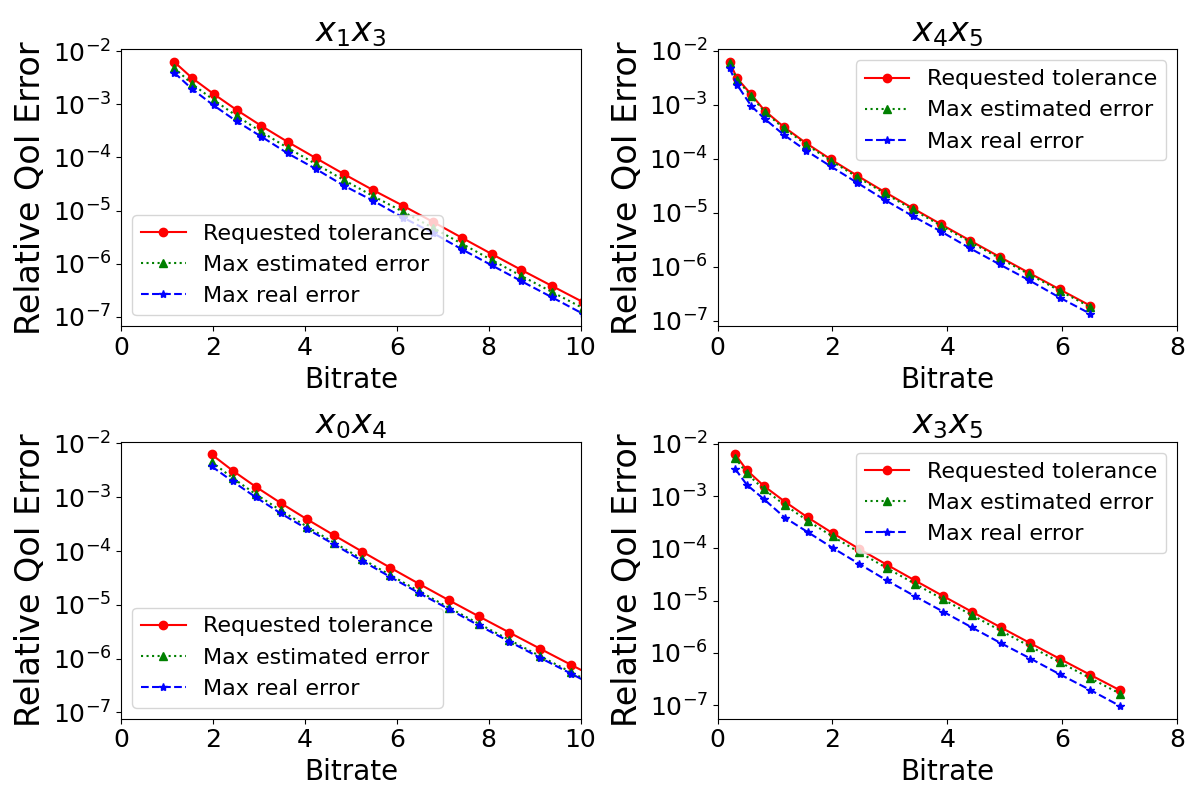}
	\centering
	\caption{Max estimated QoI errors and max actual QoI errors under given requested QoI errors of \texttt{PMGARD-HB} on S3D data.}\label{fig:qoi-s3d}
\end{figure}

We also present the results of total velocity on NYX and Hurricane data in Fig.~\ref{fig:qoi-nyx-hurricane}, as well as four examples for molar concentration multiplications on S3D data in Fig.~\ref{fig:qoi-s3d}. 
Similar trends of QoI errors in total velocity are observed in the other two datasets, which demonstrate the generality of our algorithm. 
Also, our QoI error estimator demonstrates high accuracy on S3D QoIs. This is because these QoIs only involve the multiplications of two variables, which have predictable errors in almost all cases.

\subsection{Retrieval efficiency}
We then compare the efficiency of the three progressive approaches in terms of their retrieved data sizes. 
In particular, we compare their bitrates under one requested QoI error to demonstrate generic cases; requesting progressive error bounds leads to similar results for \texttt{PSZ3-delta} and \texttt{PMGARD-HB} but may negatively impact \texttt{PSZ3}. 
Similar to the setting in Section~\ref{sec:impl}, we choose $\epsilon_i = 10^{-i}$ for $i = 1, 2, \cdots 18)$ (used 18 because some datasets such as S3D requires high precision) as the pre-set error bounds for \texttt{PSZ3} and \texttt{PSZ3-delta}. 

\begin{figure}[t]
	\includegraphics[width=\columnwidth]{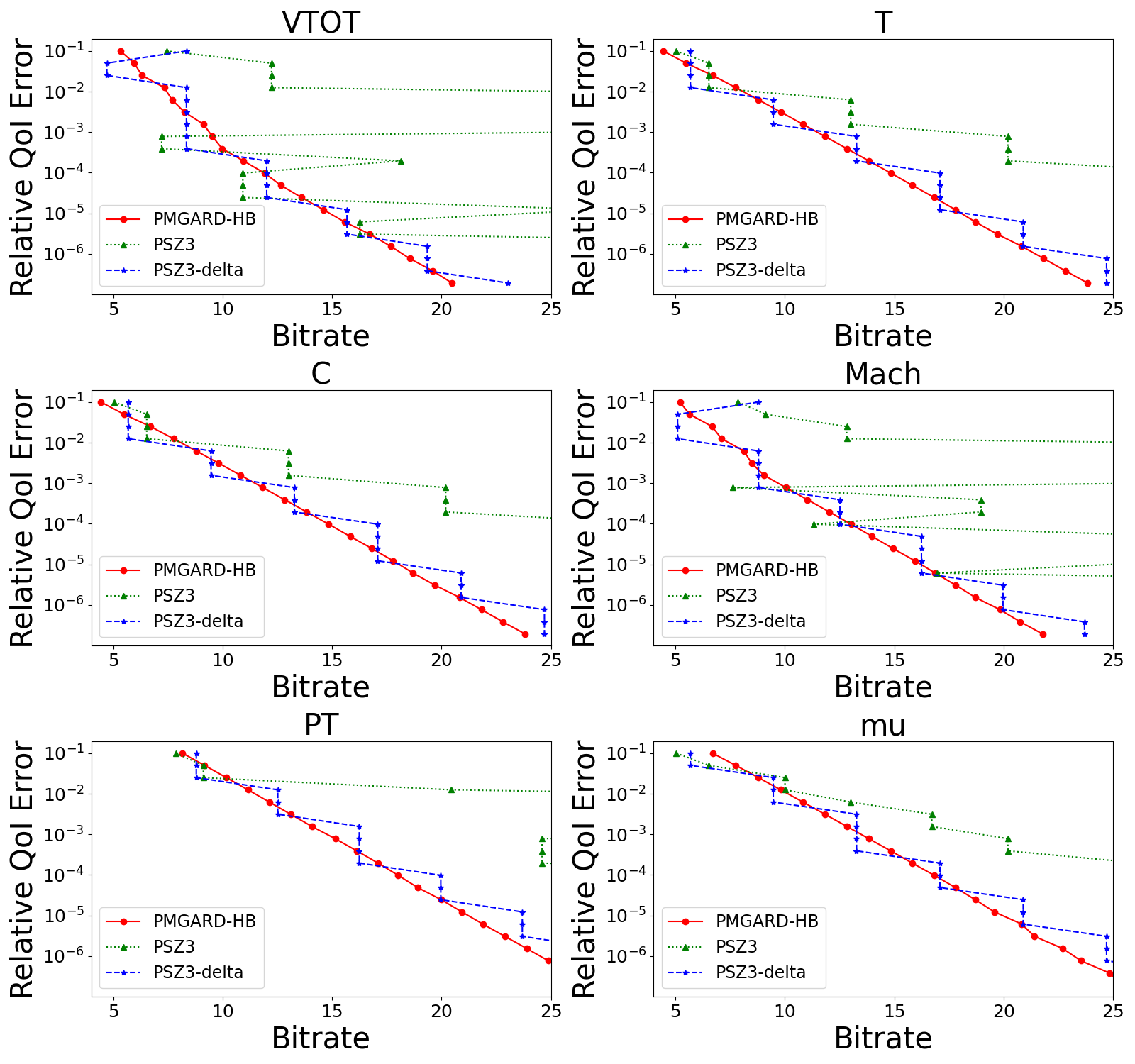}
	\centering
	\caption{Retrieval efficiency of different progressive approaches on GE-small data.}\label{fig:cmp-ge}
\end{figure}

\begin{figure}[t]
	\includegraphics[width=\columnwidth]{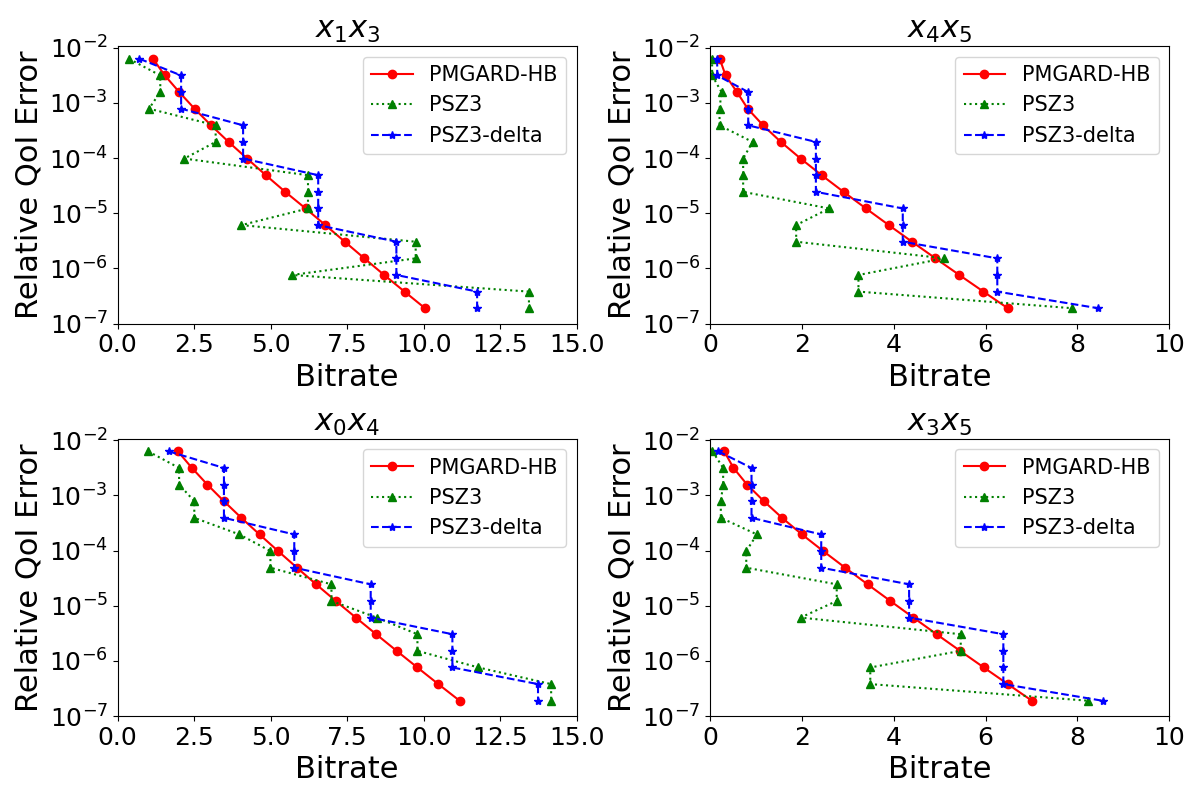}
	\centering
	\caption{Retrieval efficiency of different progressive approaches on S3D data.}\label{fig:cmp-s3d}
\end{figure}

\paragraph{Retrieved data size} 
We present the comparison of the three progressive approaches on the GE-small and S3D datasets in Fig.~\ref{fig:cmp-ge} and~\ref{fig:cmp-s3d}, respectively. The requested QoI errors are set to $\tau = 0.1*2^{-i}$ for $i=0, 1, \cdots 19$, and we omit the total velocity on NYX and Hurricane as they have similar trends to the total velocity in the GE case.  
According to these figures, \texttt{MGARD-HB} generally leads to the best bitrate among all the three methods, and it has the most steady curve; 
\texttt{PSZ3-delta} is comparable to \texttt{MGARD-HB} in most cases, but it suffers from the sudden increase of bitrate in certain ranges, which is probably caused by the use of an additional multi-precision segment; 
\texttt{PSZ3} is the least efficient in general due to the redundancy in the representation. 
In addition, it has very wild behavior when there is only a minor change on the request QoI error bound, which is probably caused by some extreme values (e.g., near-zero value in total velocity). 
Nevertheless, it performs reasonably well on S3D because of the high compressibility of the dataset and the relatively easy-to-preserve QoIs.

\paragraph{Refactoring and retrieval time} 
We present the refactoring and retrieval time of the three methods in Table~\ref{tab:ge-performance}. 
According to the table, \texttt{PMGARD-HB} has the least data refactoring time because it only needs to perform a single decomposition with bitplane encoding; in contrast, \texttt{PSZ3} and \texttt{PSZ3-delta} require the execution of the compression procedure on either original data or the residues for 18 times (equal to the number of pre-set error bounds). 
The retrieval time of the three methods is in the same order, and their differences are mainly caused by the complexity of the decompression/reconstruction algorithms and the number of iterations used to determine the proper error bound on primary data. 
\begin{table}
    \centering
    \caption{Refactor and retrieval time (seconds) of different progressive approaches on GE-small data}\label{tab:ge-performance}
    \resizebox{\columnwidth}{!}{%
    \begin{tabular}{|c|c|c|c|c|c|c|c|}
        \hline
        \multirow{2}{*}{\makecell{Compressor}} & \multirow{2}{*}{\makecell{Refactoring}}  & \multicolumn{5}{c|}{\makecell{Requested QoI error bound (VTOT)}}\\
        \cline{3-7}
         & & 1E-1 & 1E-2 & 1E-3 & 1E-4 & 1E-5 \\
        \hline
        \texttt{PMGARD-HB} & 3.30 & 0.84 & 0.95 & 1.12 & 1.34 & 1.52\\
        \hline
        \texttt{PSZ3} & 14.63 & 0.53 & 0.69 & 0.63 & 1.27 & 1.15\\
        \hline
        \texttt{PSZ3-delta} & 11.99 & 0.72 & 0.72 & 0.72 & 0.94 & 1.08\\
        \hline
    \end{tabular}
    }
    \vspace{-4mm}
\end{table}

\begin{figure}[t]
	\includegraphics[width=\columnwidth]{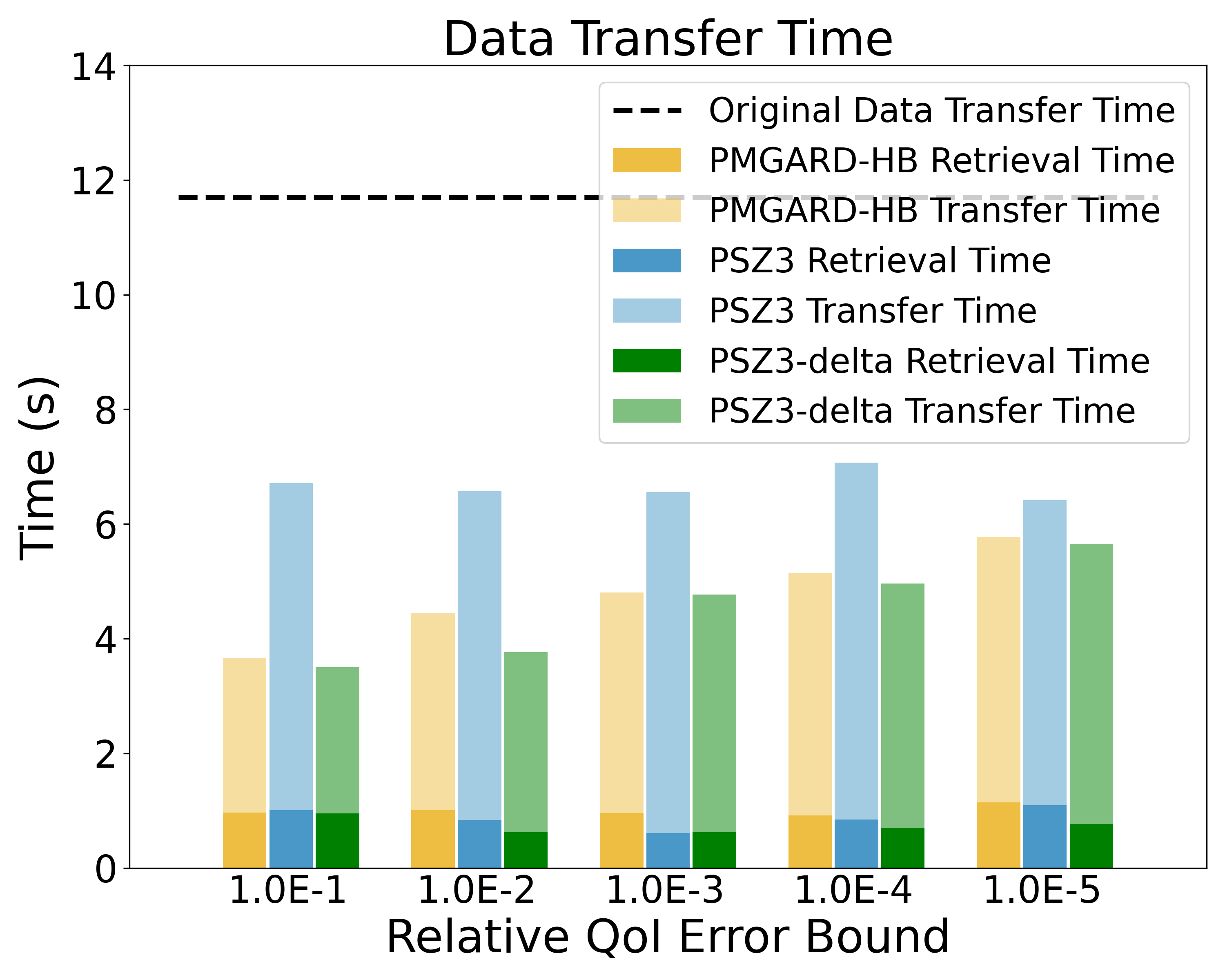}
	\centering
	\caption{Data transfer time (from MCC to Anvil via Globus) under different requested QoI error bounds with 96 cores using GE-large data. The dashed line indicates the time for transferring the original data.}\label{fig:transfer}
\end{figure}
\subsection{Remote data transfer performance}
We showcase how the proposed method can potentially improve the performance of data retrieval from remote sites using the GE-large data with VTOT as the target QoI. 
The refactored data is stored at MCC, and the data retrieval request is initiated from the Anvil supercomputer~\cite{song2022anvil} located at Purdue University. 
The experiment is performed using $96$ cores, each of which will deal with one data block in the GE-large data independently, and the data transfer is performed using the renowned data service software Globus~\cite{foster1997globus}.
The data refactoring time for \texttt{PMGARD-HB}, \texttt{PSZ3}, and \texttt{PSZ3-delta} is 2.17 seconds, 5.18 seconds, and 4.67 seconds, respectively, and the total data transfer time is depicted for remote retrieval in Fig.~\ref{fig:transfer}.  
As a baseline, the transfer time of the original data ($3$ variables, 4.67 GB in total) is roughly $11.7$ seconds, as indicated by the dashed line. 
For progressive approaches, the data transfer time includes the retrieval time, which determines the proper amount of data, and the transfer time, which is the actual time for transmitting them. 
It is observed that all the progressive approaches lead to less total data transfer time when certain QoI errors can be tolerated. 
\texttt{PMGARD-HB} and \texttt{PSZ3-delta} exhibit similar performance as they have similar sizes of reduced data in this case, but \texttt{PMGARD-HB} features a shorter data refactoring time as noted above. 
When compared with the vanilla data transfer with the original data, \texttt{PMGARD-HB} yields $2.02\times$ data transfer performance if the requested QoI error tolerance is 1E-5, because the size of the transferred data is less than 27\% of the original one. 

%% file: conclusion.tex
\section{Conclusion}\label{sec:conclusion}
In this paper, we present a progressive data retrieval framework that is able to provide QoI error control on demand. 
We derive the theory to preserve a set of derivable QoIs, and leverage them to preserve six QoIs in a computational fluid dynamics simulation from a real-world application.
Our theory is generic, and can be easily extended to preserve a wide range of QoIs that can be composited by the provided derivable QoIs. 
We further integrate three representative progressive methods into our framework and explore their efficiency in QoI preservation on five datasets from scientific applications. 
Experimental results demonstrate that the proposed framework can provide guaranteed error control on the target QoIs, which will lead to $2.02\times$ data transfer performance while ensuring a QoI error of 1E-5.
In the future, we will investigate how to extend this framework to incorporate more QoIs and progressive methods. 
We will also research how to enable tighter error controls with better efficiency.